# Transient Dynamics of a Miura-Origami Tube during Free Deployment

**Haiping Wu[1], Hongbin Fang[2,3,4,*], Lifen Chen[1], and Jian Xu[2,3,4]**

[1] Department of Aeronautics and Astronautics, Fudan University, Shanghai 200433, China

[2] Institute of AI and Robotics, Fudan University, Shanghai 200433, China

[3] Engineering Research Center of AI & Robotics, Ministry of Education, Fudan University, Shanghai 20043, China

[4] Shanghai Engineering Research Center of AI & Robotics, Fudan University, Shanghai 200433, China

[*]**Author for correspondence:** fanghongbin@fudan.edu.cn (H. Fang)

**Abstract:**

With excellent folding-induced deformability and shape reconfigurability, origami-based designs have shown great potentials in developing deployable structures. Noting that origami deployment is essentially a dynamic process, while its dynamical behaviors remain largely unexplored owing to the challenges in modeling. This research aims at advancing the state of the art of origami deployable structures by exploring the transient dynamics under free deployment, with the Miura-origami tube being selected as the object of study because it possesses relatively simple geometry, exceptional kinematic properties, and wide applications. In detail, a preliminary free deployment test is performed, which indicates that the transient oscillation in the transverse direction is nonnegligible and the tube deployment is no longer a single-degree-of-freedom (SDOF) mechanism. Based on experimental observations, four assumptions are made for modeling purposes, and a 2$N$-DOF dynamic model is established for an $N$-cell Miura-origami tube to predict the transient oscillations in both the deploying and the transverse directions. Employing the settling times and the overshoot values as the transient dynamic indexes, a comprehensive parameter study is then carried out. It reveals that both the physical and geometrical parameters will significantly affect the transient deploying dynamics, with some of the parameter dependence relationships being counter-intuitive. The results show that the relationships between the transient dynamic behaviors and the examined parameters are sometimes contradictory in the deploying and the transverse directions, suggesting the necessity of a compromise in design. Overall, this research proposes a novel method for constructing origami dynamic model. Although only the Miura-origami tube is exemplified in detail, the methodologies are promising to be applied to other origami deployable systems because the established model is believed to be able to capture the essential characteristics of origami deployment, and meanwhile, hold good processability. In addition, the comprehensive parameter analysis results could provide useful guidance for the design and optimization of Miura-origami deployable tubes with robust dynamic performance.

## 1. INTRODUCTION

Generally, a deployable structure is defined as a structure that can transform between different shapes so as to significantly change its size [1]. In the space industry, using the concept of deploying is an appealing solution to overcome the strict launch vehicle constraints. Typical space deployable structures include solar panels, antennas, booms, and masts, etc. [2]. The scope of deployable structures is not confined to space application, they have also been applied in biomedical sciences, architectures, optics, and robotics [1,3,4].

Recently, origami, the art of paper folding, has been recognized as a novel platform for developing deployable structures. While originating from art, currently, origami inclusively indicates all folding practices that transform 2-dimensional (2D) crease patterns into 3-dimensional (3D) shapes, regardless of the materials been used. Origami structures have been proved to possess unique advantages including infinite design space, excellent deformability and shape reconfigurability, flat-foldability, single-degree-of-freedom (SDOF) folding mechanism, and extraordinary folding-induced kinematic and mechanical properties. With these merits, origami opens up brand-new possibilities for developing deployable structures. On one hand, the origami principle has been utilized to interpret the underlying mechanism of natural deploying phenomena, especially, in the plant kingdom [5–7]. On the other hand, the origami-based designs have been exploited for various deployable structures in a variety of fields [8–11], including self-deployable stent graft [12], foldable robots [13–16], acoustic and optical devices [17–19], architectural structures [3,20,21], and mechanical metamaterials [22–24], etc. Particularly, origami has found its values in deployable space structures [10]. In 1985, Koryo Miura proposed a flat-foldable and rigid-foldable origami pattern, namely, the Miura-origami (for short, Miura-ori) pattern, for compactly folding solar sails and panels [25]. It has been revealed that folding of the Miura-ori pattern is an SDOF mechanism with noteworthy negative Poisson's ratio [26]. Hence, since its birth, the Miura-ori and its close relatives have become the most popular fold patterns for origami-inspired space structures [27–29]. In addition to the Miura-ori, other fold patterns that can be used for deployable space structures include the flasher patterns [30–32], the Yoshimura patterns [33], the Kresling patterns [34,35], the curved crease patterns [36], as well as various foldable cylinders, tubes, and cones [35,37,38].

With the upsurge of deep space detection and manned space missions, there is a higher request on the precision and reliability of large-scale deployable structures. Conventional research efforts that only focus on the stowed and deployed configurations would be insufficient, and there is a necessity in considering the dynamics of deployment. However, the current research level on the transient dynamics of the deployable structures lags far behind the other extensively addressed areas, including the geometric and kinematic properties [39], structural analysis of stowed and deployed configuration [40], packaging and folding techniques [25], etc. Part of the reason is that the deployable structure is designed for the space environment, and ground testing is often not a feasible option owing to the excessive size and the ground test environment. Thus, prediction of the on-orbit deploying performance heavily relies on dynamic modeling of the system. However, an accurate and processable model is always difficult to be established, because a deployable structure generally consists of numerous rigid and flexible components that are coupled together, and the physical parameters are partly unknown or difficult to be measured. McPherson and Kauffman provide a systematic review of the current work of deployable space structure dynamics [41].

For origami structures, crease pattern designs, folding kinematics, quasi-static properties, and actuation methods are the main areas of concern that have been well derived in literature [8,9,11,24,42]. Note that in practice, the origami structures may suffer from external excitations, and the folding could be a dynamic process; however, the origami dynamics research is still in its infancy. One way to build the dynamic model is to simplify the origami structure into an equivalent truss structure [43–45]. Based on this concept, nonlinear wave dynamics of Tachi-Miura polyhedron (TMP) and triangulated cylindrical origami (TCO) origami metamaterials have been explored [43,44]. Another way to derive the equation of motion is to consider the origami structure as a highly nonlinear spring [46–50], whose equivalent constitutive relation can be obtained by energetic approaches or quasi-static tension and compression tests of the origami structure. For the sake of analysis, the constitutive relation can be further approximated by polynomials. Based on this idea, Fang et al uncovered that the intrinsic bistability of the stacked Miura-ori (SMO) cell could induce complex dynamical behaviors, which have been verified by a systematic experimental investigation [46]. In addition to these two methods, pure experiment or simulation studies on the dynamics of origami structures have also been reported [51–53], which provide direct understandings on the dynamic behaviors of origami structures.

The abovementioned pioneered research on origami dynamics has been limited to the steady-state responses (e.g., wave propagations [43,44,47] or forced oscillations [46,48,54]). For origami deployable structures, the transient deploying dynamics is of greater significance in determining the performance, which, however, has not been well understood. On one hand, the transient dynamics is likely to depend strongly on the physical properties and the complex nonlinear geometries of the origami deployable structures. Analyzing such dependency relationships calls for accurate dynamic modeling, nonetheless, there is a lack of mature origami dynamic modeling methodology. On the other hand, in practice, rigid folding of the origami structure cannot be perfectly ensured, which may thus break the SDOF assumption and induce multiple-degree-of-freedom motions. Although motions other than folding is always ignorable in quasi-static folding, neglecting them in transient dynamic analyses may cause serious consequences, especially for origami space deployable structures working on orbit.

Based on the above statement, this research aims to advance the state of the art by breaking through two problems. First, how to establish an accurate and processible model for predicting the transient dynamics of a practical origami deployable structure? The two modeling techniques mentioned above, i.e., the equivalent truss [43–45] and the equivalent nonlinear spring [46–50] representations, have accomplished the first step by simplifying the complex origami structure into an analyzable dynamic system. However, admittedly, some key attributes of origami structures are missing in the simplification, such as the inertia of the origami facets; some important characteristics of folding cannot be captured by these over-simplified models, such as the additional DOF induced by non-ideal creases. Putting these factors into consideration calls for an entirely different modeling approach. Second, what are the qualitative and quantitative relationships between the transient dynamic performance and the structure parameters? Answering this question requires a systematic examination of the transient dynamics and a comprehensive parameter study, which, ultimately, could offer insights and guidance for the development of origami deployable structures with robust dynamic performance.

To address these research goals, the Miura-ori tube is selected as the object of study, and its free deployment is focused. The reasons for choosing the Miura-ori tube lie in three aspects. First, tube-like origami structures could support loads while exhibiting large folding deformations. They have wide applications in structures that call for uniaxial extensions, such as the deployable space booms and masts [10,33,37,55]. Second, the Miura-ori tube possesses the simplest geometry for analysis; its constituent unit is a degree-4 vertex cell with inherent reflectional symmetry and collinear creases. In an ideal scenario, the Miura-ori tube is rigidly foldable and with an SDOF mechanism. Although other origami tubes made of Yoshimura and Kresling patterns can exhibit richer deformations [10,45,55], they are no longer rigidly foldable. This would introduce additional complexities to the dynamic modeling and prototype fabrication. Third, practically, the rigid-foldability and SODF folding mechanism of the Miura-ori tube may be partly broken. Hence, the Miura-ori tube could still serve as a good platform for studying the transient dynamics in both the uniaxial deploying direction and the transverse direction. It is worth noting here that the aim of this research is to break through the challenges in transient dynamic modeling and parameter dependence relationship, exemplified by the Miura-ori tube, rather than putting forward a new deployable structure design. Nevertheless, the proposed modeling technique and the developed dynamic analysis method are promising to be applied to other non-soft origami deployable structures.

The rest of this paper is organized as follows. Section II introduces the kinematic model of the Miura-ori tube and demonstrates the results of a preliminary free deployment test. Insights from the preliminary experiment could provide useful hints for dynamic modeling and analysis. Section III details the dynamic modeling technique, including modeling assumptions and the equation of motion. Based on this, a comprehensive parameter study is carried out in Section IV. The effects of the physical and geometric parameters on the deploying dynamic performance, characterized by the settling times and the overshoot values, are provided and interpreted. Summary and conclusions of this research, as well as a heuristic discussion, are presented in Section V.

## II. RIGID-FOLDING KINEMATICS AND PRELIMINARY EXPERIMENTS

In this section, rigid-folding kinematics of the Miura-ori tube is firstly introduced. Then a 6-cell prototype is fabricated, and preliminary free deployment tests are carried out to get insights into the fundamental characteristics of the transient dynamics.

## A. Rigid-folding kinematics of the stacked Miura-ori cell

Figure 1(a) shows a Miura-ori tube consisting of six stacked-Miura origami (SMO) cells, each SMO cell is stacked by two Miura-ori units with kinematic constraints (Fig. 1(b)). In detail, the geometries of a Miura-ori unit can be characterized by three parameters, namely, the lengths of two adjacent crease lines $a_k$ and $b_k$, and the sector angle $\gamma_k$ between them, where $k = \text{A}, \text{B}$, denoting the bottom unit A and the top unit B, respectively. Without loss of generality, the bottom unit A is assumed to have the shorter crease length, i.e., $a_\text{A} < a_\text{B}$. With the rigid-origami assumption, folding of the Miura-ori unit is a single-degree-of-freedom (SDOF) mechanism, which can be described by the folding angle $\theta_k$ ($k = \text{A}, \text{B}$), defined as the dihedral angle between the facet and the reference *x-y* plane (Figure 1(c)). To ensure kinematic compatibility between the two units so that they remain connected during the whole folding process, the following constraints have to be satisfied [26,56]

$$b_\text{A} = b_\text{B} = b, \quad a_\text{A} \cos\gamma_\text{A} = a_\text{B} \cos\gamma_\text{B}. \tag{1}$$

When stacked together, folding of the two units are no longer independent of each other; rather, the folding angles $\theta_k$ ($k = \text{A}, \text{B}$) are constrained by

$$\cos\theta_\text{A} \tan\gamma_\text{A} = \cos\theta_\text{B} \tan\gamma_\text{B}. \tag{2}$$

Equation (2) indicates a non-unique relation between the folding angles $\theta_\text{A}$ and $\theta_\text{B}$. When $\theta_\text{A} \in [-\pi/2, \pi/2]$, $\theta_\text{B}$ remains non-negative; in other words, for a given positive value of $\theta_\text{B}$, $\theta_A$ could take two values with the same magnitude but opposite signs. In what follows, $\theta_A$ is employed to describe the folded configuration of the SMO cell. When $\theta_A > 0$, the bottom unit A nests into the top unit B, and the configuration is denoted as "nested-in"; when $\theta_A < 0$, the bottom unit A bulges out of the top unit B, and the configuration is denoted as "bulged-out."

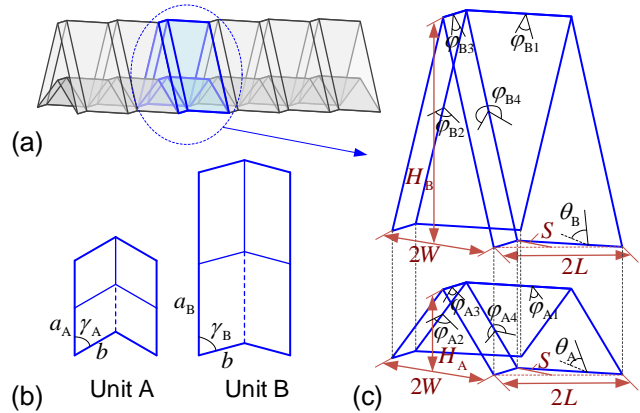

**Fig. 1.** Geometries of the stacked Miura-ori cell. (a) A Miura-ori tube composed of six identical SMO cells, where a constituent cell is highlighted. (b) Planar crease patterns of the constituent Miura-ori units A and B, where the internal solid and dashed lines indicate the "mountain" and "valley" creases, respectively. (c) Dihedral angles and external dimensions of an SMO cell, showing in an exploded view.

The folded configuration of a Miura-ori unit can also be characterized by the dihedral angles between adjacent facets $\varphi_{ki}$ ($k = A, B$; $i = 1, 2, 3, 4$) (Figure 1(c)), which are functions of the folding angles $\theta_k$. Based on spherical trigonometry, we have [46]

$$\sin\frac{\varphi_{k2}}{2} = \frac{\cos\theta_k}{\sqrt{1-\sin^2\theta_k \sin^2\gamma_k}}, \quad (3)$$
$$\varphi_{k1} = \varphi_{k3} = \pi - 2\theta_k, \quad \varphi_{k4} = 2\pi - \varphi_{k2}.$$

Here, for the nested-in configuration ($\theta_A > 0$), we assign $\varphi_{A2} \in (0, \pi)$; and for the bulged-out configuration ($\theta_A < 0$), we assign $\varphi_{A2} \in (\pi, 2\pi)$. At the connecting creases between the two constituent Miura-ori units, the dihedral angle

$$\varphi_C = \theta_B - \theta_A. \quad (4)$$

Based on the above geometry analysis, the external dimensions of an SMO cell can be derived [26]

$$L = \frac{b\sin\gamma_A \cos\theta_A}{\sqrt{1-\sin^2\theta_A \sin^2\gamma_A}}, \quad W = a_A\sqrt{1-\sin^2\theta_A \sin^2\gamma_A}$$
$$S = \frac{b\cos\gamma_A}{\sqrt{1-\sin^2\theta_A \sin^2\gamma_A}}, \quad H_k = a_k \sin\theta_k \sin\gamma_k \quad (k = A, B). \quad (5)$$

### B. Prototype design and fabrication

To get a basic understanding of the transient dynamic behavior of the Miura-ori tube during free deployment, preliminary experiments are carried out. To this end, a Miura-ori tube is prototyped, with the geometry parameters listed in Table 1. The fabrication procedures are demonstrated in Figure 2. In detail, the origami facets are laser cut individually from 0.25-mm thick stainless steel sheets, and 0.1-mm thick adhesive-back plastic films (ultrahigh molecular weight (UHMW) polyethylene) are used to connect the facets into two separate 6-cell Miura-ori sheets (top sheet and bottom sheet, Fig. 2(a)). To assign the origami structure with appropriate folding stiffness, based on our past experiences [46], three 0.1-mm thick pre-bent spring-steel stripes are pasted at the creases with length $a_B$ in the top Miura-ori sheet (Fig. 2(b)). The pre-bent angle of the spring-steel stripes is about $85.3°$, corresponding to a stress-free configuration of the SMO structure with $\theta_A^0 = 60°$. Then the top sheet and the bottom sheet are stacked and connected via adhesive-back plastic films, generating a 6-cell tube prototype. Figure 2(c) shows the photos of the prototype (in the nested-in configuration) from two different angles of view. Considering that the constituent SMO cell is mirror-symmetric, the tube is fundamentally made up of 12 half SMO cells ("half-cell" for short in what follows). A single half-cell is indicated in Fig. 2(c) by a dashed rectangle.

**TABLE 1.** Geometries of the Miura-ori tube prototype

| Parameters | $a_A$[mm] | $a_B$[mm] | $b$[mm] | $\gamma_A$[deg] | $\gamma_B$[deg] |
|---|---|---|---|---|---|
| Values | 38.1 | 73.6 | 38.1 | 60 | 75 |

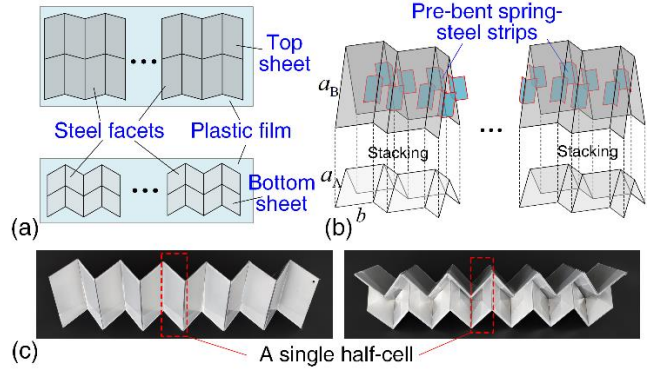

**Fig. 2.** Fabrication of the Miura-ori tube prototypes. (a) Connecting laser-cut steel facets into top and bottom Miura-ori sheets via adhesive-back plastic film. (b) Stacking the top sheet and bottom sheet into a tube. Pre-bent spring-steel stripes are pasted at the creases with length $a_B$. (c) Photos of the 6-cell prototype (in the nested-in configuration) from two angles of view, where a single half-cell is denoted.

### C. Setup for free deployment test

Figure 3(a) and 3(b) show the sketch and the photo of the experimental setup. The Miura-ori tube prototype (in the nested-in configuration) is fixed to a customized acrylic panel at one end via foldable steel sheets and movable screw rods (Fig. 3(c)). In detail, on the acrylic panel, a groove and two sliding chutes are designed based on the rigid-folding kinematic of the SMO cell (Fig. 3(d)). At point $H_1$, a pair of foldable steel sheets are used to connect the prototype with the panel, with one sheet inserted into the panel through the groove, and the other sheet connected with the prototype at the crease via adhesive-back plastic film. At points $H_2$ and $H_3$, screw rods are utilized to connect the prototype with the panel. For each screw rod, one end is fixed with the prototype along the crease via adhesive-back plastic films, and the other end is free to move along the sliding chute. The chute trajectory corresponding to points $H_2$ and $H_3$ is an arc and a horizontal line segment, respectively; they are determined based on

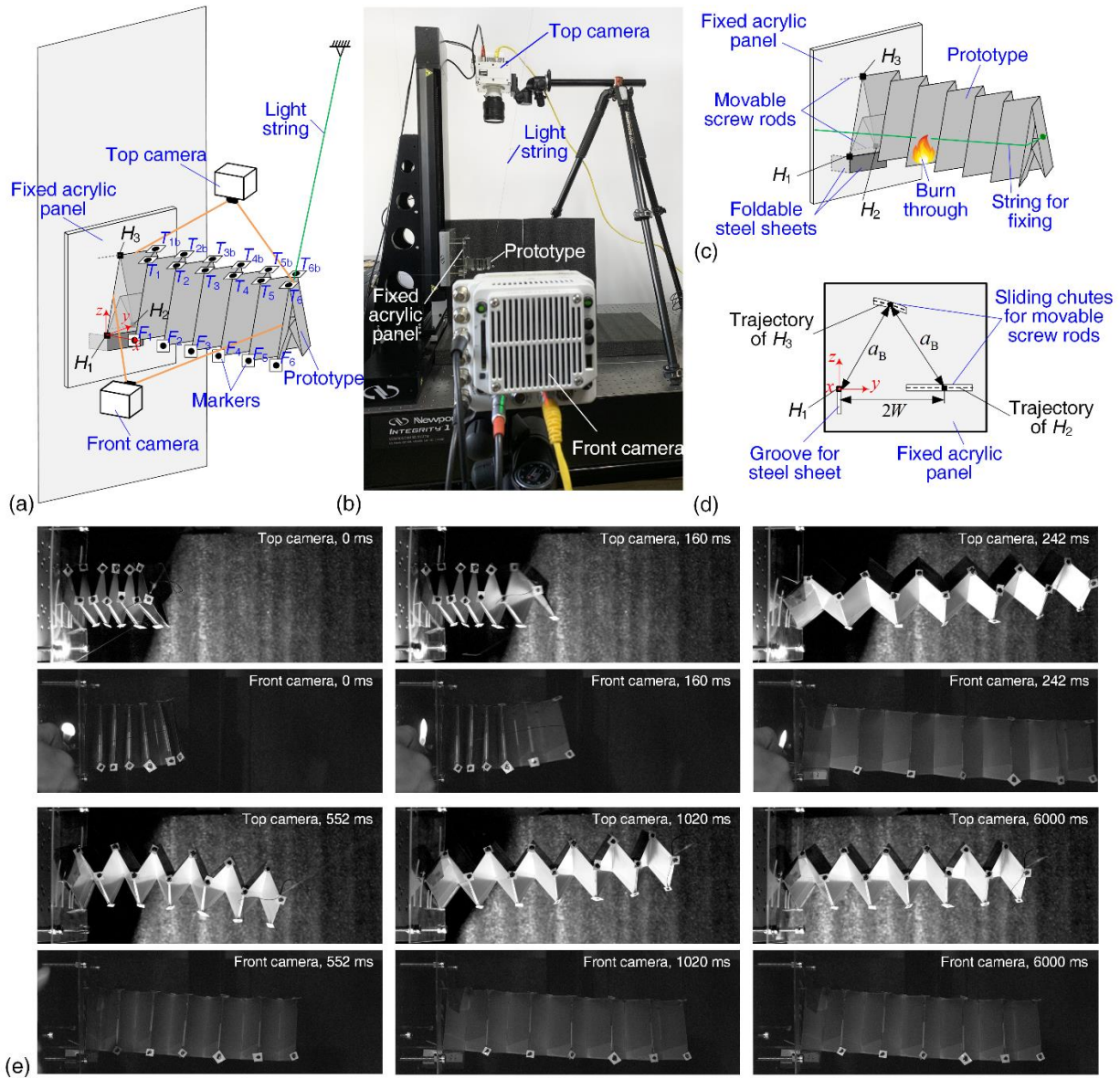

**Fig. 3.** Preliminary experiment on the free deployment of a 6-cell Miura-ori tube. (a) and (b) show the sketch and photo of the experimental setup, respectively. Displacements of the markers on the prototype are measured by the top and the front high-speed cameras. (c) demonstrates how the tube prototype connects with the acrylic panel through points $H_1$, $H_2$, and $H_3$, as well as how the tube is released from a compressed configuration via burning through the string for fixing. The design of the customized acrylic panel is illustrated in (d). For description purposes, a Cartesian coordinate system is set with the origin locating at point $H_1$. (e) displays the images obtained from the two high-speed cameras at six key time instants during free deployment.

$$\left|H_1H_3\right| = a_B, \quad \left|H_2H_3\right| = a_B, \quad \left|H_1H_2\right| = 2W, \tag{6}$$

where $W$ has been derived in Eq. (5). Based on the above setup, folding of the related facets is not affected. However, the displacements of point $H_1$ in all directions are completely constrained, points $H_2$ and $H_3$ are only allowed to move along the chutes. For easy descriptions, a global Cartesian coordinate system is set with the origin locating at point $H_1$, the x-axis pointing along the deploying direction, and the y-axis pointing along $H_1H_2$. Hence, we have $x_{H_1} = y_{H_1} = z_{H_1} = 0$, $x_{H_2} = 0$, $z_{H_2} = 0$, and $x_{H_3} = 0$.

To examine the displacements of the constituent SMO cells in the *x*, *y*, and *z* directions, six markers (denoted by $F_1$ to $F_6$) are pasted at the outermost vertices in the front of the prototype; and another 12 markers (denoted by $T_1$ to $T_6$, and $T_{1b}$ to $T_{6b}$) are pasted at the top vertices of the prototype (Fig. 3(a)). In the experiments, the tube prototype firstly stays at a compressed configuration with the help of a string for fixing. The free end of the tube is suspended from the ceiling by a light string to reduce the effects of gravity. By setting a fire and burning through the string, the compressed prototype is released to perform a free deployment. Two high-speed cameras (Phantom® VEO 640), with one arranged on the top of the prototype, and the other arranged in front of the prototype, are used to record the displacements of the markers (Fig. 3(a) and 3(b)). In detail, displacements of the constituent cells in the *x* and *y* directions can be read from the markers $T_1$ to $T_6$ via the top camera; displacements in the *z*-direction can be read from the markers $F_1$ to $F_6$ via the front camera. The sampling rate of the cameras is set to be 1000 fps (frame per second).

### D. Preliminary experiments and understandings

Based on the abovementioned setup, free deployment tests are carried out on the prototype. The high-speed camera record frames for no less than six seconds. Figure 3(e) display the images obtained from the high-speed cameras at six key time instants, namely, the initial instant (0 ms), the instant when the string is burnt through (160 ms), the first time when the tube is deployed to the largest length (242 ms), the instant when the tube exhibits the largest displacement in the negative *y*-direction (552 ms), the instant when the tube exhibits the largest displacement in the positive *y*-direction (1020 ms), and the final instant when the cameras stop recording (6000ms). Figure 4 displays the measured displacement-time histories of the six constituent SMO cells in the *x*, *y*, and *z* directions, respectively. The following phenomena are observed and interpreted.

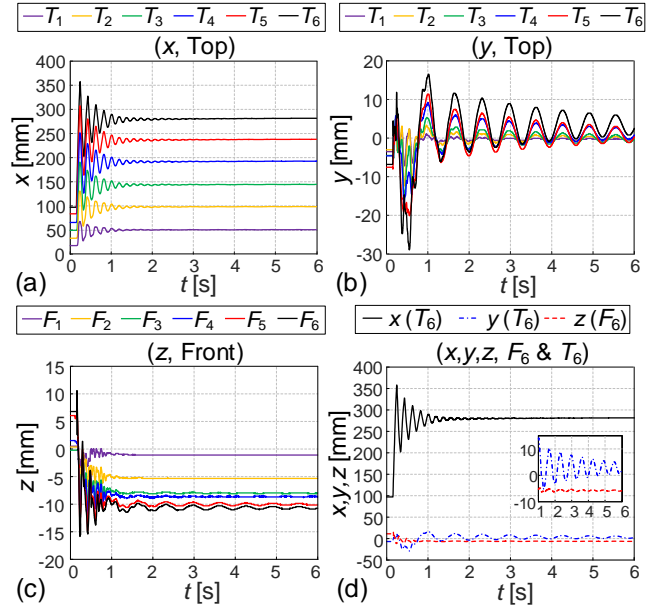

**Fig. 4.** Transient response of the Miura-ori tube prototype in a free deployment test, recorded by high-speed cameras. (a)~(c) Displacement-time histories of the constituent cells in the $x$, $y$, and $z$ directions, respectively (measured from the markers $T_i$ and $F_i$ ($i=1,...6$)). (e) For the cell at the free end, the displacement responses in the $x$, $y$, and $z$ directions (measured from the markers $T_6$ and $F_6$) are compared; the insect shows an enlarged view of the displacement-time histories in the $y$ and $z$ directions.

(i) In addition to the expected transient vibrations along the deploying direction (i.e., the $x$-direction), the tube exhibits significant transverse vibrations in the $y$-direction. Transverse vibrations in the $z$-direction are also detected, however, overall, the vibration amplitudes are much lower than those in the $y$-direction.

(ii) In the $x$, $y$, and $z$ directions, the transient vibrations of the constituent cells are synchronized in general. However, their vibration amplitudes are distinctly different. The farther away from the fixed end, the larger the amplitude is. Particularly, the cell at the free end (measured from markers $T_6$ and $F_6$, Fig. 4(a)~(c)) always exhibits the largest vibration amplitudes. To have a clear comparison, the corresponding displacement-time histories are extracted and displayed in Fig. 4(d).

(iii) Focusing on the cell at the free end (characterized by markers $T_6$ and $F_6$), we notice that the transient vibration amplitude in the $y$-direction is much more significant than that in the $z$ direction. During the initial releasing stage (within 800 ms), due to the imperfect pre-compressed configuration and the non-ideal fixed-end connection, the vibration is relatively irregular, with the largest peak-peak amplitude being 40.82 mm in the

*y*-direction and 26.45 mm in the *z*-direction. After that, the oscillation in the *z*-direction decays fast; around 1000 ms, the largest peak-peak amplitude is only 2.63 mm in the *z*-direction, which is about 7.6 times lower than that in the *y*-direction (22.62 mm).

Note that with the rigid-folding assumption, folding of the whole Miura-ori tube is an SDOF mechanism, and transverse displacements of the tube cannot occur. However, in our experiment, significant transverse vibrations in the *y*-direction have been observed. This is because, in a real tube prototype, the creases made of plastic films cannot be considered as ideal rotational hinges anymore; rather, they would exhibit non-negligible deformations, inducing rotational deviations between adjacent facets. In such a scenario, an SDOF rigid-folding dynamic model is insufficient to predict the transient dynamic behavior of a real origami tube, and it becomes a necessity to set up a new dynamic model by considering the non-ideal creases.

## III. DYNAMIC MODELING

Fundamentally, the Miura-ori tube is a rigid-flexible coupled system, and establishing its dynamic model is generally a challenging task. Fortunately, the preliminary free deployment test provides some useful hints to simplify dynamic modeling.

### A. Model assumptions

A generic *N*-cell Miura-ori tube consists of 2*N* half cells. For ease of descriptions, '*i, j*' is used to denote each vertex. The first number '*i*' indicates the *i*-th half-cell ($i = 0,...,2N$), with '*i*=0' referring to the fixed end and '*i*=2*N*' referring to the free end; the second number '*j* ($j = 1,...,4$)' indicates the vertices at the connection between the *i*-th and (*i*+1)-th cells, arranged in a counterclockwise direction. As an example, Fig. 5(a) shows the vertex numbers in two adjacent SMO cells (i.e., four half-cells). In what follows, such a numbering method will continue to be used, including the subscriptions.

Based on the observations we obtained from the preliminary experiment, four assumptions are made for dynamic modeling purposes.

***Assumption 1:*** *Rigid facets and ideal creases within each half-cell.* During deployment, folding of the constituent half-cells is still dominated by rigid-folding. Hence, when developing the dynamic model, the origami facets are assumed to be rigid, and the creases within each half-cell are assumed to be ideal hinges (denoted by solid lines in Fig. 5(a)). With this assumption, the rigid-folding kinematics of each half-cell can be strictly followed.

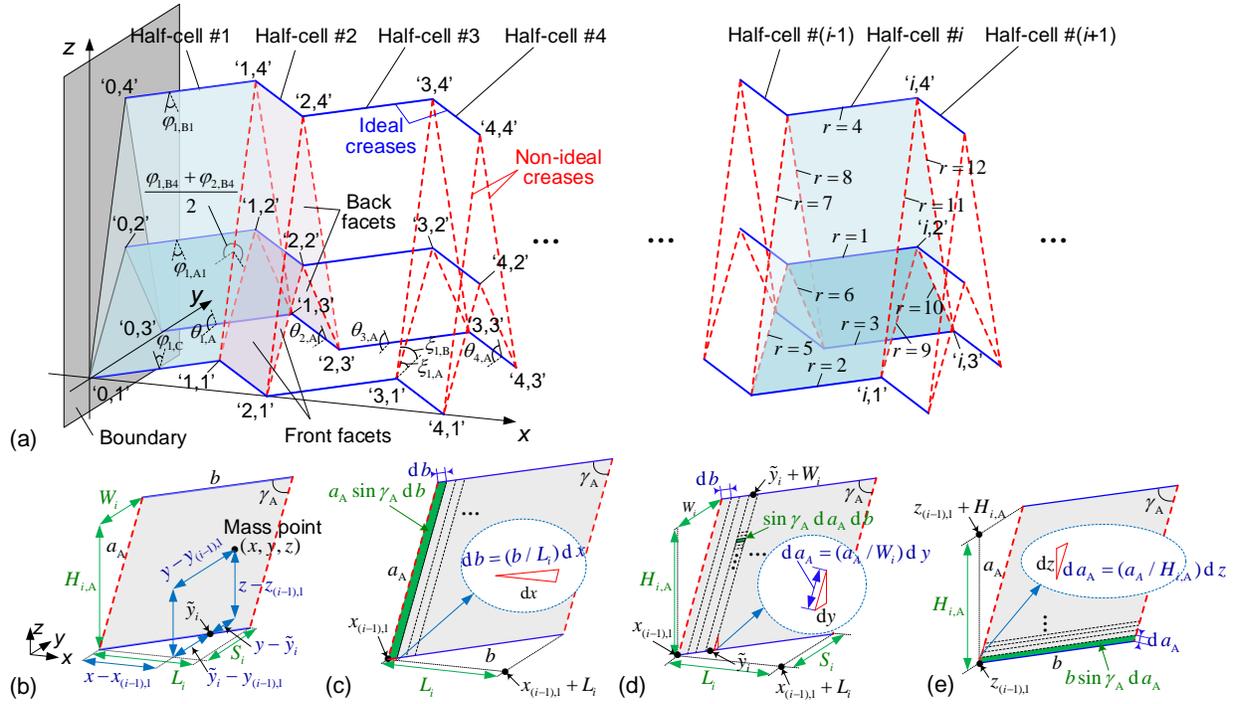

**Fig. 5**. Illustration of the dynamic modeling process. (a) A part of the tube, which consists of 7 half-cells, namely, half-cells #1 to #4, half-cells #($i$-1), #$i$, and #($i$+1). The ideal creases within each half-cell are denoted by solid lines, and the non-ideal creases shared by adjacent half-cells are denoted by dashed lines. Some angles used for modeling are exemplified, and the crease sequence numbers $r$ ($r$=1,…,12) in the $i$-th half-cell are denoted. (b) As an example, a mass point locating on a facet of the $i$-th half-cell is indicated. The dimensions used for expressing the speeds of the mass point are denoted. (c)~(f) demonstrate how surface integrations are performed in the $x$, $y$, and $z$ directions, respectively.

***Assumption 2:*** *Non-ideal creases between adjacent half-cells and additional elastic constraints.* The creases shared by adjacent half cells are not ideal (denoted by dashed lines in Fig. 5(a)). Hence, folding motions of adjacent half-cells can deviate. Nevertheless, the non-ideal creases would still impose strong constraints between adjacent half-cells. To describe such constraints, an elastic stiffness $k^*$ is additionally introduced to the shared creases to constrain the folding deviation between adjacent half-cells. Its contribution to the dynamic model will be introduced in detail in Section IIIB2.

***Assumption 3:*** *Boundary conditions.* As is introduced in Fig.3(c), at the fixed end, vertex '0,1' is fully constrained in all the three directions (i.e., $x_{0,1} = y_{0,1} = z_{0,1} = 0$); vertex '0,3' is only allowed to translate in the $y$-direction (i.e., $x_{0,3} = z_{0,3} = 0$); vertices '0,2' and '0,4' are allowed to translate within the $y$-$o$-$z$ plane (i.e., $x_{0,2} = 0$ and $x_{0,4} = 0$). At the free end of the tube, all vertices are free to move within the space.

***Assumption 4:*** *Partial displacement continuity.* Displacements are asked to be continuous at the vertices '$i,1$' ($i = 1,...,2N-1$) in all the three directions. However, the displacement continuity conditions at vertices '$i,2$', '$i,3$', and '$i,4$' are not fully satisfied due to the possible deformations of the shared creases between adjacent half-cells. At vertex '$i,3$', displacements are asked to be continuous in both the $x$ and $z$ directions. At vertices '$i,2$' and '$i,4$', displacements are asked to be continuous only in the $x$-direction.

Based on Assumptions 1, 3, and 4, displacements of the tube in the $z$-direction will be restrained, while displacements in the $y$-direction are admissible, which agree with the observation from the preliminary experiments that the vibration in the $z$-direction is much weaker than that in the $y$-direction. Based on Assumption 2, to describe the dynamics of the tube, the DOF has to be increased to the number of constituent half-cells (i.e., $2N$). Moreover, based on Assumptions 1, 3, and 4, for a fixed value of '$i$' ($i = 0,...,2N$), the vertices '$i,j$' ($j = 1,...,4$) will be restricted on a plane parallel to the $y$-$o$-$z$ plane.

### B. Dynamical modeling

To describe the dynamics of the Miura-ori tube consisting of $2N$ half-cells, $2N$ generalized coordinates are needed, which are prescribed as the folding angles of the half-cells, i.e., $\boldsymbol{\theta} = (\theta_{1,A}, \theta_{2,A}, ..., \theta_{2N,A})^T$ (see examples in Fig. 5(a)). Based on the above assumptions, the governing equations of the tube can be constructed via the Lagrange equation for the general case

$$\frac{d}{dt}\left(\frac{\partial(T-V)}{\partial \dot{\theta}_{i,A}}\right) - \frac{\partial(T-V)}{\partial \theta_{i,A}} = Q_i, \tag{7}$$

where $T$ and $V$ denote the kinetic and potential energy of the tube, respectively, $Q_i$ denotes the nonconservative generalized force associated with the generalized coordinate $\theta_i$.

#### 1. Kinetic energy

The total kinetic energy $T$ of the tube can be expressed as the sum of the kinetic energies of the constituent half-cells, i.e.,

$$T = \sum_{i=1}^{2N} T_i. \tag{8}$$

The kinetic energy of the $i$-th half-cell $T_i$ can be further expressed as the sum of the kinetic energies in the $x$, $y$, and $z$ directions,

$$T_i = \sum_{k=A,B} \sum_{s=x,y,z} T_{i,ks}. \tag{9}$$

For $T_{i,ks}$, the subscript $i$ ($i=1,...,2N$) denotes the $i$-th half-cell, $k$ ($k=A,B$) denotes the bottom sheet A or the top sheet B, and $s$ ($s=x,y,z$) denotes the direction.

Due to the intrinsic reflection symmetry of the half-cell, the velocities of the mass points on the front and back facets are identical in the $x$ and $z$ directions but are different in the $y$-direction. For the mass point locating on the front and back facets (Fig. 5(a)), the velocity in the $y$-direction is denoted by $v_{i,ky}$ and $v'_{i,ky}$, respectively. By considering the geometric relations (Fig. 5(b)), the velocities of a mass point (with coordinate $(x,y,z)$) on the facet of the $i$-th half-cell can be determined, which yields

$$v_{i,kx} = \dot{x}_{(i-1),1} + \frac{x - x_{(i-1),1}}{L_i} \dot{L}_i,$$

$$v_{i,ky} = \dot{y}_{(i-1),1} + \frac{x - x_{(i-1),1}}{L_i} \dot{S}_i + \frac{y - \tilde{y}_i}{W_i} \dot{W}_i,$$

$$v'_{i,ky} = \left(\dot{y}_{(i-1),1} + \dot{W}_i\right) + \frac{x - x_{(i-1),1}}{L_i} \dot{S}_i + \frac{y - (\tilde{y}_i + W_i)}{W_i} \dot{W}_i, \quad (10)$$

$$v_{i,kz} = \dot{z}_{(i-1),1} + \frac{z - z_{(i-1),1}}{H_{i,k}} \dot{H}_{i,k},$$

where $\tilde{y}_i = y_{(i-1),1} + S_i(x - x_{(i-1),1})/L_i$. In these equations, $s_{i,j}$ ($s=x,y,z; i=1,...,2N; j=1,...,4$) denotes the absolute position of the vertex '$i,j$' in the $i$-th half-cell; $v_{i,ks}$ ($i=1,...,2N$; $k=A,B$; $s=x,y,z$) denotes the velocity of a mass point on the facet of the bottom sheet ($k=A$) or the top sheet ($k=B$) of the $i$-th half-cell in the $x$, $y$, or $z$-direction. $L_i$, $W_i$, $H_{i,k}$, and $S_i$ are the external dimensions of the $i$-th half-cell, given in Eq. (5).

Denoting the areal density of the facet by $\rho$, the kinetic energy $T_{i,ks}$ can be derived via areal integral. Figure 5(c)~5(e) exemplifies how the surface integrations are performed in the three directions.

$$T_{i,kx} = 2 \times \frac{1}{2} \int_{x_{(i-1),1}}^{x_{(i-1),1}+L_i} \rho \left( \frac{a_k b \sin \gamma_k}{L_i} \right) v_{i,kx}^2 dx,$$

$$T_{i,ky} = \frac{1}{2} \int_{x_{(i-1),1}}^{x_{(i-1),1}+L_i} \rho \left( \frac{a_k b \sin \gamma_k}{L_i W_i} \right) \left[ \int_{\tilde{y}_i}^{\tilde{y}_i+W_i} v_{i,ky}^2 dy + \int_{\tilde{y}_i+W_i}^{\tilde{y}_i+2W_i} (v'_{i,ky})^2 dy \right] dx, \quad (11)$$

$$T_{i,kz} = 2 \times \frac{1}{2} \int_{z_{(i-1),1}}^{z_{(i-1),1}+H_{i,k}} \rho \left( \frac{a_k b \sin \gamma_k}{H_{i,k}} \right) v_{i,kz}^2 dz.$$

After integration, they yield

$$T_{i,kx} = \rho a_k b \sin \gamma_k \left( \frac{1}{3} \dot{L}_i^2 + \dot{x}_{(i-1),1}^2 + \dot{L}_i \dot{x}_{(i-1),1} \right),$$

$$T_{i,ky} = \rho a_k b \sin \gamma_k \left( \frac{1}{3} \dot{S}_i^2 + \frac{4}{3} \dot{W}_i^2 + \dot{y}_{(i-1),1}^2 + 2\dot{W}_i \dot{y}_{(i-1),1} + \dot{S}_i \dot{y}_{(i-1),1} + \dot{W}_i \dot{S}_i \right), \quad (12)$$

$$T_{i,kz} = \rho a_k b \sin \gamma_k \left( \frac{1}{3} \dot{H}_{i,k}^2 + \dot{z}_{(i-1),1}^2 + \dot{H}_{i,k} \dot{z}_{(i-1),1} \right).$$

Substituting Eq. (12) into Eq. (8) and (9), the total kinetic energy $T$ of the tube can be obtained.

## 2. *Potential energy*

The total potential energy $V$ of the tube has three origins: the potential energy from the ideal hinge-like creases (of length $b$) within each half-cell $V_b$, the potential energy from the shared creases between adjacent half-cells $V_S$, and the potential energy due to the additionally introduced stiffness at the connecting creases $V_\theta$, i.e.,

$$V = V_b + V_S + V_\theta. \quad (13)$$

Generally, the creases of the tube are assigned with certain rotational stiffness. For the ideal hinge-like creases within each half-cell, the rotational stiffness per unit length is $k$ (Fig. 5(a)). Hence, the potential energy $V_b$ can be derived as

$$V_b = \sum_{i=1}^{2N} \frac{1}{2} kb \left[ (\varphi_{i,A1} - \varphi_{i,A1}^0)^2 + (\varphi_{i,B1} - \varphi_{i,B1}^0)^2 + 2(\varphi_{i,C} - \varphi_{i,C}^0)^2 \right], \quad (14)$$

where the angle $\varphi_{i,k1}$ ($i = 1,...,2N; k = A, B$) denotes the dihedral angle between the two adjacent facets of the bottom or top sheet in the $i$-th half-cell, and the angle $\varphi_{i,C}$ denotes the dihedral angle at the connecting creases between the top and bottom sheet in the $i$-th half-cell (see examples in Fig. 5(a)). They can be analytically obtained by substituting $\theta_{i,k}$ for $\theta_k$ in Eq. (3) and (4). The superscript '0' represents the angle at the stress-free configuration with $\theta_A^0$.

For the creases that are shared by adjacent half-cells in the bottom sheet and the top sheet, the rotational stiffness per unit length is $k$ and $\mu k$, respectively (Fig. 5(a)). The constant $\mu > 1$ characterizes the stiffness difference. In our experimental prototype, this difference is induced by the embedded pre-bent spring-steel belts. Corresponding to the shared creases between the ($i-1$)-th and the $i$-th half-cells, the dihedral angles can be expressed as $(\varphi_{(i-1),k2} + \varphi_{i,k2})/2$ and $(\varphi_{(i-1),k4} + \varphi_{i,k4})/2$, where $\varphi_{i,k2}$ and $\varphi_{i,k4}$ ($k = A, B$) can be obtained by substituting $\theta_{i,k}$ for $\theta_k$ in Eq. (3) (see an example in Fig. 5(a)). The potential energy $V_S$, therefore, includes two parts, namely, the energies from the shared creases in the bottom sheet A ($V_{SA}$) and the top sheet B ($V_{SB}$). Considering that $\varphi_{i,k4} = 2\pi - \varphi_{i,k2}$ always holds, we have

$$V_S = V_{SA} + V_{SB},$$

$$V_{SA} = 2 \sum_{i=1}^{2N-1} \frac{1}{2} k a_A \left( \frac{\varphi_{i,A2} + \varphi_{(i+1),A2}}{2} - \frac{\varphi^0_{i,A2} + \varphi^0_{(i+1),A2}}{2} \right)^2, \quad (15)$$

$$V_{SB} = 2 \sum_{i=1}^{2N-1} \frac{1}{2} \mu k a_B \left( \frac{\varphi_{i,B2} + \varphi_{(i+1),B2}}{2} - \frac{\varphi^0_{i,B2} + \varphi^0_{(i+1),B2}}{2} \right)^2.$$

Based on Assumption 2, the shared creases would also constrain the folding deviation between adjacent half-cells. This is achieved by the additional stiffness $k^*$, which applies to the angular deviations of the shared creases between adjacent half-cells, i.e.,

$$V_\theta = 2 \sum_{i=1}^{2N-1} \frac{1}{2} k^* \left[ (\xi_{i,A} - \xi_{(i+1),A})^2 + (\xi_{i,B} - \xi_{(i+1),B})^2 \right]. \quad (16)$$

Here, the angle $\xi_{i,k}$ ($i=1,...,2N$; $k=A,B$) is defined as the angle between the crease and the x-o-y plane (see an example in Fig.5(a)), which has the form

$$\cos \xi_{i,A} = \sqrt{1 - \sin^2 \gamma_A \sin^2 \theta_{i,A}},$$

$$\cos \xi_{i,B} = \frac{\cos \gamma_B}{\cos \gamma_A} \sqrt{1 - \sin^2 \gamma_A \sin^2 \theta_{i,A}}. \quad (17)$$

Generally, $k^*$ is much larger than $k$, we define $\lambda = k^*/k$. The larger $\lambda$ is, the stronger the constraint; particularly, if $\lambda \to \infty$, the crease angular deviation is completely prohibited, which would degenerate the system into an SDOF rigid-folding dynamic model. Substituting Eq. (14)~(16) into Eq. (13), the total potential energy $V$ of the tube can be obtained.

### 3. Nonconservative generalized force

In addition to providing rotational stiffness, generally, the creases would also dampen the deploying of the tube. Here, viscous damping is assumed to each crease, the damping coefficient per unit length is $c_0$. Based on the principle of virtual work, the nonconservative generalized force associated with the generalized coordinate $\theta_{i,A}$ is

$$Q_i = \sum_{r=1}^{12} \left( -c_0 l_{i,r} \frac{d\psi_{i,r}}{dt} \right) \frac{\partial \psi_{i,r}}{\partial \theta_{i,A}}, \quad (18)$$

where $r = 1,...,12$ indicates the 12 crease lines in the $i$-th half-cell (see notations in Fig. 5(a)), $l_{i,r}$ denotes the length of the $r$-th crease, and $\psi_{i,r}$ are the corresponding dihedral angles. Hence, $c_0 l_{i,r} (d\psi_{i,r}/dt)$ is the viscous damping force associated with the $r$-th crease in the $i$-th half-cell. Detailed derivations of the nonconservative generalized forces are given in the supplementary information.

## 4. *Equation of motion*

Substituting the kinetic energy, potential energy, and the generalized forces of the tube into the Lagrange equation (i.e., Eq. (7)) and performing necessary simplification, the equation of motion of the Miura-ori tube can be obtained, which can be expressed as the following matrix form

$$\mathbf{J}\ddot{\boldsymbol{\theta}} + \mathbf{C}\dot{\boldsymbol{\theta}} + \mathbf{G}\dot{\boldsymbol{\theta}}^2 + \mathbf{F}_V = \mathbf{0}, \tag{19}$$

where $\mathbf{J}\ddot{\boldsymbol{\theta}}$ represents the inertial forces, $\mathbf{C}\dot{\boldsymbol{\theta}}$ represents the viscous damping forces, $\mathbf{G}\dot{\boldsymbol{\theta}}^2$ represents the Coriolis and centrifugal forces, their dimensions are $2N \times 2N$, and $\mathbf{F}_V$ is the restoring force vector. Detailed derivations of the equation of motion are given in the supplementary information.

### C. Transient dynamic responses: an example

Based on Eq. (19), we are then able to analyze the transient dynamics of the tube during free deployment. Here, we first examine an example to demonstrate the effectiveness of the model. Specifically, a 6-cell (i.e., $N = 6$) Miura-ori tube (in the nested-in configuration) is studied. The geometric parameters of the tube are the same as the prototype (listed in Table 1), and the physical parameters are listed in Table 2. The tube is firstly pre-compressed to an initial configuration where all constituent half-cells share the same folding angle, i.e., $\theta_{i,\mathrm{A}} = \theta_{\mathrm{Initial}}$ ($i = 1,...,12$). Then the pre-compressed tube is released, and the transient deploying dynamics can be obtained by numerically solving Eq. (19) via the 4$^{\mathrm{th}}$ order Runge-Kutta method. Vibrations of the constituent SMO cells can be characterized by vertices '$2i,1$' ($i = 1,...,6$), because displacements on vertices '$2i,1$' ($i = 1,...,6$) are continuous in all directions (based on Assumption 4). Figure 6(a) and 6(b) displays the displacement-time histories of the vertices '$2i,1$' ($i = 1,...,6$) in the *x* and *y* directions, respectively. It reveals that the tube exhibits significant vibrations in both the deploying *x*-direction and the transverse *y*-direction, and the vertex '12,1' at the free end always vibrates with the largest amplitudes, which fully agree with the experimental observation. Hence, in what follows, the vertex '$2N,1$' (within the half-cell at the free end) will serve as the representative of the transient dynamics. For this example, the displacement-time histories of the vertex '12,1' are extracted and displayed in Fig. 6(c) and 6(d).

To quantify the transient dynamic performance of the tube, two indexes are examined in this research: the overshoot value and the settling time of vertex '12,1' (Fig. 6(c) and 6(d)). Specifically, the overshoot value refers to the maximum value that exceeds its final, steady-state value, which can be either positive or negative; the settling time is defined as the time required for the output to reach and remain within a given error band. Here, the error bands are prescribed as ±2.5 mm in the deploying *x*-direction and ±0.5 mm in the transverse *y*-direction, respectively.

**TABLE 2. Physical parameters of the Miura-ori tube**

| Parameters | Values | Parameters | Values |
|---|---|---|---|
| $k$ | 20 mN/rad | $\rho$ | $1.95 \times 10^{-6}$ kg/mm² |
| $\mu$ | 10 | $\theta_A^0$ | 60° |
| $\lambda$ | 20000 | $\theta_{Initial}$ | 86.4° |
| $c_0$ | 5 kg/(s·rad) | | |

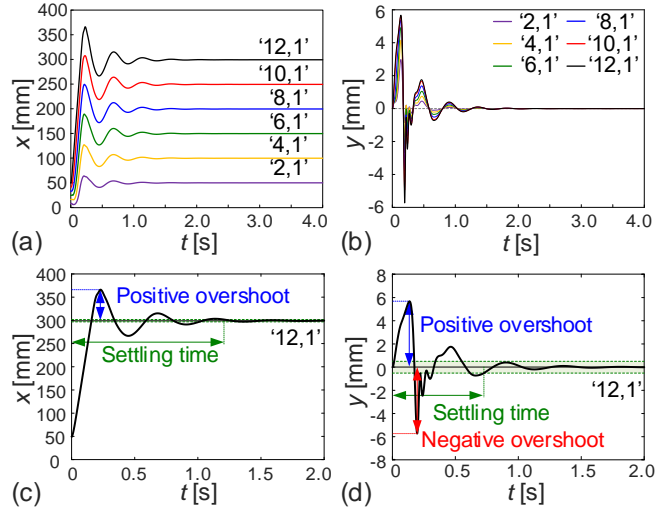

**Fig. 6**. Free deployment transient dynamics of a 6-cell Miura-ori tube. (a) and (b) show the displacement-time histories of the vertices '$2i,1$' ($i = 1,...6$) in the deploying *x*-direction and the transverse *y*-direction, respectively. (c) and (d) show the displacement-time histories of the vertex '12,1' in the *x* and *y* directions, respectively. The examined indexes, namely, the overshoot values and the settling time, are denoted. The shaded bars indicate the prescribed error bands.

## IV. EFFECTS OF PHYSICAL AND GEOMETRIC PARAMETERS ON THE TRANSIENT DYNAMIC PERFORMANCE

In this section, we study how the physical and geometrical parameters affect the transient dynamic performance of the Miura-ori tube under free deployment. To this end, a 6-cell ($N = 6$)

Miura-ori tube (in the nested-in configuration) is studied. In what follows, without repeating, two key indexes are examined, namely, the settling time ($T_{s,x}$ and $T_{s,y}$ in the $x$ and $y$ directions, respectively) and the overshoot values ($\hat{x}$ in the $x$-direction, $\hat{y}_+$ and $\hat{y}_-$ in the $y$-direction) of vertex '12,1'. We believe that an in-depth understanding of the parameter dependence relationships could provide useful guidance to the development of origami deployable tubes.

### A. Physical parameters

This subsection aims to uncover the relationship between the physical parameters and the transient dynamic performance under free deployment. The adjustable physical parameters are the crease stiffness $k$, the stiffness ratios $\mu$ and $\lambda$, and the viscous damping coefficient $c_0$.

#### 1. Effects of the additional stiffness $k^*$

The stiffness $k^*$, or equivalently, the ratio $\lambda = k^*/k$, is used to characterize the perfectness of the creases shared by adjacent half-cells; in other words, $k^*$ or $\lambda$ characterizes the strength of the constraints between adjacent half-cells. The larger the value of $k^*$ or $\lambda$ is, the more ideal the creases are, and the stronger the constraints applied to the folding deviation between adjacent half-cells. Figure 7 displays how the settling times and the overshoot values evolve with $\lambda$. The other parameters remain the same as those listed in Tables 1 and 2.

Figure 7(a) reveals that the stiffness $k^*$ has little effects on the settling time in the deploying $x$-direction. By increasing the ratio $\lambda$ from 2 to $2\times10^8$, the settling time $T_{s,x}$ slightly declines from 1.22 s to 1.13 s. However, in the transverse $y$-direction, the effect on the settling time $T_{s,y}$ is significant. With the increase of $\lambda$ from 2 to $2\times10^8$, $T_{s,y}$ decreases notably in the overall trend, from 2.10 s to zero. Specifically, between points $P_1$ ($\lambda = 2$) and $P_2$ ($\lambda = 89.3$), $T_{s,y}$ declines quickly with $\lambda$, following which, there is an upturn in $T_{s,y}$ between points $P_2$ and $P_3$ ($\lambda = 8.9\times10^2$). These phenomena can be interpreted in terms of the equivalent damping ratio. When $\lambda$ is relatively small, the system is overdamped with the equivalent damping ratio larger than 1 (e.g., point $P_1$). With the rise of $\lambda$, the equivalent damping ratio falls off, which diminishes the settling time $T_{s,y}$. At a certain value, the equivalent damping ratio would drop below 1, thus switching the system from overdamped (e.g., point $P_1$) to underdamped (e.g., point $P_3$) and inducing an elevation of $T_{s,y}$. Further enlarging $\lambda$, although the equivalent damping ratio keeps decreasing, the stiffness $k^*$ becomes large enough to effectively constrain the folding deviations between adjacent half-cells and suppress the transverse vibrations, which as a result, substantially reduces the settling time $T_{s,y}$ again (e.g., at points $P_4$ ($\lambda = 1.2\times10^4$) and $P_5$ ($\lambda = 2\times10^7$), $T_{s,y}$ has been decreased to 0.75s and 0.15s, respectively). At very large values, say, point $P_6$ with $\lambda = 2 \times 10^8$, the settling time converges to zero, indicating that the transverse vibration in the *y*-direction vanishes. This is because the stiffness $k^*$ is sufficiently high to constrain the folding motions of adjacent half-cells and make them consistent. In such a scenario, the MDOF dynamic model degenerates into an SDOF rigid-folding dynamic model. The above analyses are verified via the displacement-time histories of the vertex '12,1' corresponding to the six points $P_1$ to $P_6$, shown in Fig. 7(c) and 7(d).

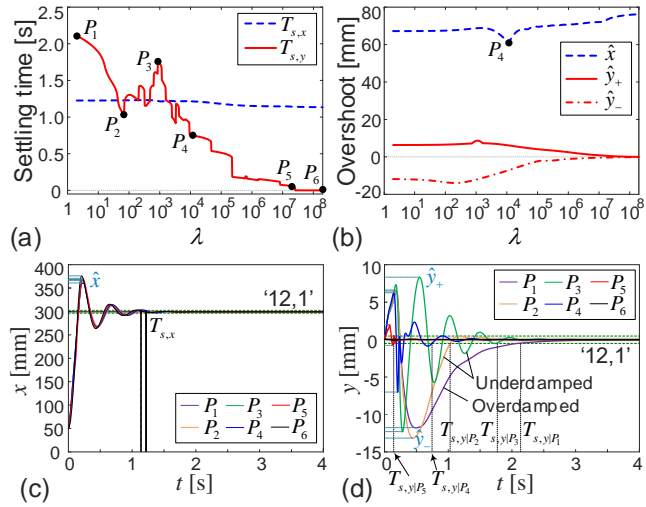

**Fig. 7.** Effects of the additional stiffness $k^*$ on the transient dynamics of the Miura-ori tube under free deployment. (a) Effects on the settling times in the *x* and *y* directions. Six points $P_1$ to $P_6$ are picked out to exemplify the evolution. (b) Effects on the overshoot values in the *x* and *y* directions. (c) and (d) show the displacement-time histories of the vertex '12,1' corresponding to the six points in the *x* and *y* directions, respectively. The settling times and the overshoot values are indicated by dashed lines. Note that $\lambda$ has little effects in the *x*-direction, while in the *y*-direction the effects are more significant.

The stiffness $k^*$ also affects the overshoot values. In the *x*-direction, except for an abrupt drop of $\hat{x}$ at point $P_4$, the overshoot value $\hat{x}$ rises slowly with $\lambda$ in the overall trend. On the contrary, in the *y*-direction, corresponding to the evolution from the overdamped scenario to the underdamped scenario, the overshoot value $|\hat{y}_-|$ experiences a gentle climb, while $\hat{y}_+$ remains steady. After that, since the transverse vibrations are effectively suppressed, the overshoot values ($\hat{y}_+$ and $|\hat{y}_-|$) decline significantly. When $\lambda = 2 \times 10^8$, both $\hat{y}_+$ and $\hat{y}_-$ becomes zero, suggesting the model degeneration again.

In sum, Fig. 7 suggests that it is necessary to consider the dynamic effects of the additional stiffness $k^*$ (or the stiffness ratio $\lambda$) when designing and modeling origami deployable tubes. Particularly, putting the equivalent damping ratio into consideration could prevent performance deterioration in transverse vibrations. Generally, with a relatively large stiffness $k^*$ (or the stiffness ratio $\lambda$), although there is a small growth of the overshoot value in the folding direction, vibrations in the transverse direction can be effectively restrained.

## 2. Effects of the crease stiffness $k$ and the stiffness ratio $\mu$

The crease stiffness $k$ and the stiffness ratio $\mu$ are two important parameters that determine the mechanical properties of the origami tube. Figure 8 displays their effects on the transient dynamics, with $k$ varying from 5 to 50 mN/rad and $\mu$ varying from 1 to 50. The other parameters are the same as those listed in Tables 1 and 2.

Figure 8(a) reveals that in the deploying direction, the settling time $T_{s,x}$ undulates with the increase of $k$ and $\mu$. Overall, $T_{s,x}$ is larger when $k$ or $\mu$ takes relatively smaller values. Particularly, the peak of $T_{s,x}$ locates in the domain where both $k$ and $\mu$ are relatively small. Contrary to expectations, the trough of $T_{s,x}$ does not occur in the domain where both $k$ and $\mu$ tend to the maximum; rather, it locates within an arc-shaped stripe where $k$ and $\mu$ take moderately large values. Unlike the settling time, the overshoot value $\hat{x}$ monotonically increasing with $k$ and $\mu$ in the overall trend (Fig. 8(b)). The peak and bottom are achieved when $k$ and $\mu$ reaches the maxima and minima, respectively.

In the transverse direction, the trend of the settling time goes into a sharp reverse (Fig. 8(c)). Overall, the peak of $T_{s,y}$ is reached when $k$ and $\mu$ tend to the maxima, and the valley of $T_{s,y}$ locates within an arc-shaped stripe where $k$ and $\mu$ are relatively small. The overall shoot values ($\hat{y}_+$ and $\hat{y}_-$) shear the same overall trend with that of $\hat{x}$, i.e., they monotonically increasing with $k$ and $\mu$ (Fig. 8(d) and 8(e)). The trend of $\hat{y}_-$ exhibits better monotonicity than $\hat{y}_+$.

The above analyses on the effects of $k$ and $\mu$ provide useful guidelines for assigning stiffness. On one hand, in terms of the settling time, a compromise is necessary because $T_{s,x}$ and $T_{s,y}$ exhibits the opposite trend. For example, at point $Q_1$ ($\mu=6$, $k=6$), the settling time $T_{s,x}(Q_1)=1.39$ s, which is longer than that at point $Q_2$ ($\mu=40$, $k=40$) with $T_{s,x}(Q_2)=1.18$ s (Fig. 8(f)). However, in the y-direction, point $Q_2$ outstrips point $Q_1$, with $T_{s,y}(Q_1)=0.78$ s, and $T_{s,y}(Q_2)=1.09$ s (Fig. 8(g)). On the other hand, in terms of the overshoot, taking relatively small values of $k$ and $\mu$ is always beneficial. For example, the overshoot values ($\hat{x}$, $\hat{y}_+$, and $\hat{y}_-$) corresponding to point $Q_1$ are always much lower than those corresponding to point $Q_2$ (Fig. 8(f) and 8(g)).

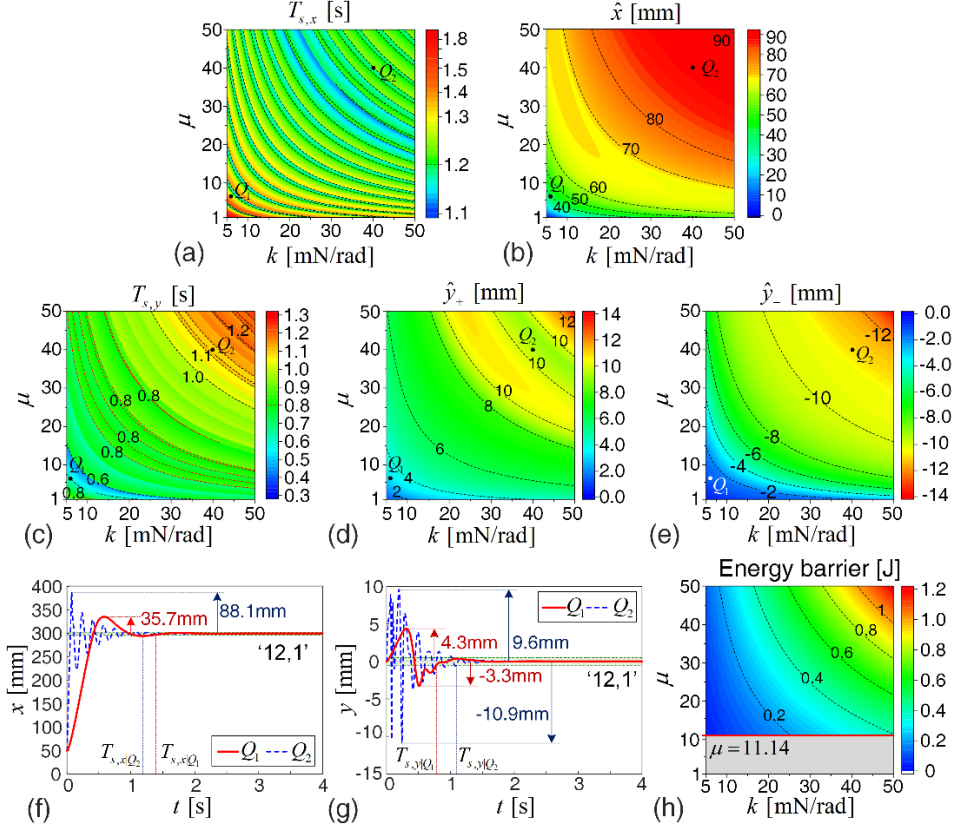

**Fig. 8.** Effects of the stiffness $k$ and ratio $\mu$ on the transient dynamics of the Miura-ori tube under free deployment. (a) and (b) show the effects on the settling time and the overshoot value in the *x*-direction, respectively. (c)~(e) show the effects on the settling time and the overshoot values in the *y*-direction, respectively. Two points $Q_1$ and $Q_2$ are picked out to exemplify the evolution. (f) and (g) show the displacement-time histories of the vertex '12,1' corresponding to the two points in the *x* and *y* directions, respectively. The settling times and the overshoot values are indicated. (g) shows the effects on the overall stability profile of the tube. Below the critical value $\mu = 11.14$ (shaded), the system is mono-stable; in the bistable region ($\mu > 11.14$), contours of the energy barrier level are plotted. The energy barrier level is defined as the energy difference between the unstable configuration and the stress-free configuration.

In addition to the two indexes, we note that the transient behaviors corresponding to points $Q_1$ and $Q_2$ are significantly different in terms of their oscillation frequencies and decaying trends (Fig. 8(f) and 8(g)). This can be interpreted from the perspective of equivalent stiffness and equivalent damping ratios. For example, a comparison of the displacement-time histories suggests that the equivalent damping ratios corresponding to point $Q_1$ in both the *x* and *y* directions are much larger than those corresponding to point $Q_2$.

It is also worth point out that changing the values of $k$ and $\mu$ would fundamentally alter the stability characteristics of the tube. As is shown in Fig. 8(h), when $\mu > 11.14$, the tube possesses different degrees of bistability, which has pros and cons. On one hand, bistability could endow the tube with the unique capability of fast and accurate configuration switch; on the other hand, bistability is a global strong nonlinearity that may trigger complex dynamics [46]. When designing origami tubular deployable structures, this issue deserves attention.

### 3. *Effects of the damping coefficient* $c_0$

Figure 9(a) and 9(b) show the effects of the damping coefficient $c_0$ on the settling times ($T_{s,x}$, $T_{s,y}$) and the overshoot values ($\hat{x}$, $\hat{y}_+$, and $\hat{y}_-$). The other parameters are the same as those listed in Tables 1 and 2. In the deploying direction, with the increase of $c_0$ from 1 to 10 $\text{kg} \cdot \text{mm/s}$, the settling time $T_{s,x}$ decay sharply to a tiny value; after that, the downtrend becomes gentle, and $T_{s,x}$ reaches the valley 0.27 s at $c_0 = 23.02 \text{ kg} \cdot \text{mm/s}$. Such a reduction of $T_{s,x}$ can be attributed to the increase of the equivalent damping ratio. If $c_0$ continues growing, surprisingly, $T_{s,x}$ starts to climb again. We find that the equivalent damping ratio will reach the critical value at point $R_3$ with $c_0 = 24.92 \text{ kg} \cdot \text{mm/s}$. After exceeding this value, the system works in the overdamped scenario, and a further increase of the damping ratio would significantly slow the deploying process, thus, enlarging $T_{s,x}$. Particularly, when $c_0$ takes a very large value, say, $c_0 = 100 \text{ kg} \cdot \text{mm/s}$ at point $R_5$, $T_{s,x}$ is increased to 2.37 s. Figure 9(c) shows the displacement-time histories of the vertex '12,1' in the *x*-direction corresponding to five characteristic values of $c_0$, from which, evolution of the system from the underdamped scenario ($R_1$ and $R_2$) to the critically damped point ($R_3$), and to the overdamped scenario ($R_4$ and $R_5$), accompanying with the abovementioned variation trend of $T_{s,x}$, can be clearly observed.

In the transverse direction, being similar with $T_{s,x}$, the settling time $T_{s,y}$ drops fast when $c_0$ increases from 1 to 10 $\text{kg} \cdot \text{mm/s}$. However, unlike in the *x*-direction that $T_{s,x}$ stops to increase, $T_{s,y}$ will stabilize at a very low level. Although some small rise is detected, $T_{s,y}$ cannot gain a large boost any more even $c_0$ takes a very large value. The variation trend of $T_{s,y}$ can be verified by the displacement-time histories of the vertex '12,1' in the *y*-direction corresponding to five characteristic values of $c_0$ ($R_1$ to $R_5$), shown in Fig. 9(d).

The effects of $c_0$ on the overshoot values are much simpler (Fig. 9(b))). In both the deploying and the transverse directions, the overshoot values ($\hat{x}$, $\hat{y}_+$ and $\hat{y}_-$) decreases monotonously with the increase of $c_0$. Specifically, $\hat{x}$ experiences a significant reduction from 80 mm to 0 by increasing $c_0$ from 1 to 50 $\text{kg} \cdot \text{mm/s}$. In the transverse direction, although $\hat{y}_+$ and $\hat{y}_-$ also decrease with $c_0$, the reductions are limited, with $\hat{y}_+$ dropping from 8.4 mm to 0.85 mm, and $\hat{y}_-$ dropping from -10.4 mm to 0. It can be expected that they will converge to zero when $c_0$ takes a very large value.

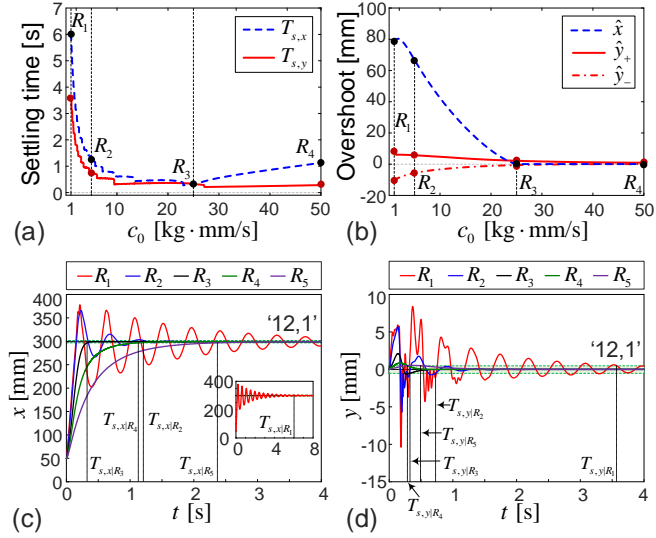

**Fig.9.** Effects of the damping coefficient $c_0$ on the transient dynamics of the Miura-ori tube under free deployment. (a) Effects on the settling times. (b) Effects on the overshoot values. To demonstrate the evolution trends, five points are picked out, namely, $R_1$ ($c_0 = 1$ kg·mm/s), $R_2$ ($c_0 = 5$ kg·mm/s), $R_3$ ($c_0 = 24.92$ kg·mm/s), $R_4$ ($c_0 = 50$ kg·mm/s), and $R_5$ ($c_0 = 100$ kg·mm/s). The corresponding displacement-time histories of the vertex '12,1' in the $x$ and $y$ directions are displayed in (c) and (d), respectively.

Figure 9 suggests that an overlarge value of $c_0$ is not always helpful. To achieve optimum transient performance during free deployment, $c_0$ should be such chosen that the equivalent damping ratio reaches the critically damped value.

## B. Geometric parameters

This subsection studies the effects of the design geometries on the transient dynamics. For the Miura-ori tube, the tailorable geometric factors are the crease length ratio $a_A / b$, the sector angles $\gamma_k$ ($k = A, B$), the number of constituent SMO cells $N$, and the tube configurations.

### 1. Effects of the crease length ratio $a_A / b$ and the sector angle $\gamma_A$

The effects of the geometry parameters on the settling times and overshoot values are shown in Fig. 10. In detail, the crease length $a_A$ and the sector angle $\gamma_A$ are assumed to be variable ($a_A / b \in [0.5, 2.0]$, $\gamma_A \in [\pi/4, 5\pi/12]$), while the crease length $b$ and the sector angle $\gamma_B$ are fixed, which are listed in Table 1. The physical parameters are the same as those listed in Table 2. Figure 10(a) reveals that in the deploying direction, the settling time $T_{s,x}$ climbs when $\gamma_A \to \pi/4$ and $a_A/b \to 2.0$. For the overshoot value $\hat{x}$, it stays at relatively high levels when $\gamma_A < \pi/3$. Particularly, $\hat{x}$ reaches the peak at $\gamma_A = 0.925, a_A/b = 1.36$, and reaches the minimum when $\gamma_A = 5\pi/12, a_A/b = 0.5$ (Fig. 10(b)).

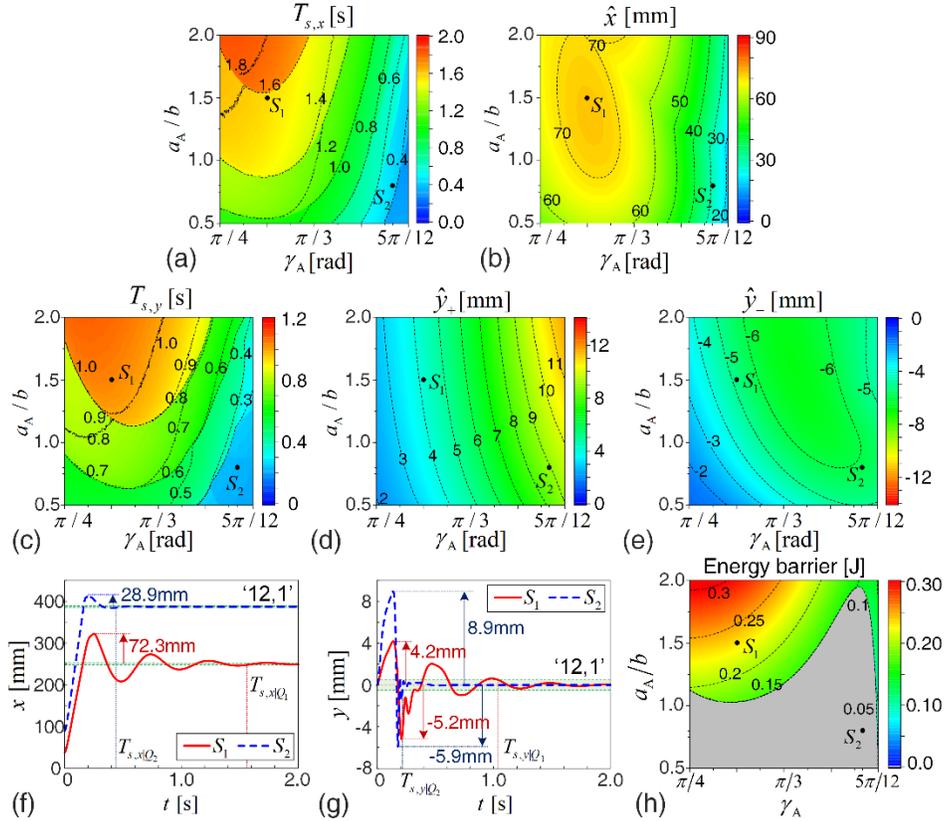

**Fig. 10.** Effects of the geometric parameters $a_A/b$ and $\gamma_A$ on the transient dynamics of the Miura-ori tube under free deployment. (a) and (b) show the effects on the settling time and the overshoot value in the x-direction, respectively. (c)~(e) show the effects on the settling time and the overshoot values in the y-direction, respectively. Two points $S_1$ and $S_2$ are picked out to exemplify the evolution. (f) and (g) show the displacement-time histories of the vertex '12,1' corresponding to the two points in the x and y directions, respectively. The settling times and the overshoot values are indicated. (g) shows the effects on the overall stability profile of the tube. The monostable region is denoted by the shade; in the bistable region, contours of the energy barrier level are plotted.

In the transverse direction, the settling time $T_{s,y}$ shares a similar trend with $T_{s,x}$ (Fig. 10(c)). However, the contours of the overshoot values ($\hat{y}_+$ and $\hat{y}_-$) are significantly different with $\hat{x}$ (Fig. 10(d), 10(e)). Overall, $\hat{y}_+$ and $\hat{y}_-$ remain at low levels when $\gamma_A < \pi/3$. Particularly, $\hat{y}_+$ peaks at $\gamma_A = 5\pi/12$, $a_A/b = 2.0$ and minimizes at $\gamma_A = \pi/3$, $a_A/b = 0.5$.

The above-concluded evolution trends of the settling times and the overshoot values are verified via the displacement-time histories of vertex '12,1' corresponding to points $S_1$ ($\gamma_A = 0.92$, $a_A/b = 1.5$) and point $S_2$ ($\gamma_A = 1.3$, $a_A/b = 0.8$) in the $x$ and $y$ directions, respectively (Fig. 10(f), 10(g)). Obviously, in terms of the settling time, point $S_2$ behaves better because $T_{s,x|S_2} < T_{s,x|S_1}$, and $T_{s,y|S_2} < T_{s,y|S_1}$. However, in terms of the overshoot values, design compromise is needed because $\hat{x}_{S_2} < \hat{x}_{S_1}$, while $\hat{y}_{+|S_2} > \hat{y}_{+|S_1}$, $\hat{y}_{-|S_2} > \hat{y}_{-|S_1}$. We also remark here that the geometric parameters $a_A/b$ and $\gamma_A$ would also affect the stability profile. Figure 10(h) shows that both monostable and bistable profiles can be achieved, which needs extra attention during the design.

### 2. Effects of the number of constituent SMO cells $N$

We also study how the number of constituent SMO cells affects the transient behavior of the tube. With $N$ SMO cells in a tube, the degree of freedom becomes $2N$. Figures 11(a) and 11(b) show the effects on the settling times ($T_{s,x}$ and $T_{s,y}$) and the overshoot values ($\hat{x}$, $\hat{y}_+$, and $\hat{y}_-$), with $N$ varying from 1 to 8. As expected, both the settling times and the overshoot values are positively correlated with $N$. However, it is worth noting that the indexes in the deploying direction are more sensitive to the increase of $N$, which calls for additional attention in the design.

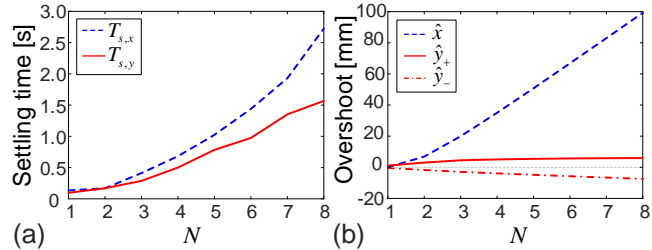

**Fig. 11.** Effects of the number of constituent cells on the transient dynamics of the Miura-ori tube under free deployment. (a) Settling times. (b) Overshoot values.

### 3. Effects of the tube configurations

Note that the Miura-ori tube possesses two topologically different configurations, the nested-in and the bulged-out configurations. It is then necessary to comment on whether the transient dynamic performance will have a significant change when the tube switches its configuration. To this end, a nested-in tube and a bulged-out tube under free deployment are investigated. For the bulged-out configuration, the stress-free angle is changed to $\theta_A^0 = -60°$, and the initial configuration is changed to $\theta_{\text{Initial}} = -86.4°$. The other geometric and physical parameters are the same as those listed in Tables 1 and 2. Corresponding to the two configurations, the displacement-time histories of the vertex '12,1' in the $x$ and $y$ directions are displayed in Fig.

12(a) and 12(b), respectively. It reveals that qualitatively, they make no big difference in the overall trend. Quantitatively, the settling time and the overshoot values corresponding to the nested-in configuration are a little higher than those corresponding to the bulged-out configuration, which can be interpreted in terms of the tangent stiffness [57]. As a result, aiming at different requirements, both configurations are applicable, with small differences in free deployment performance.

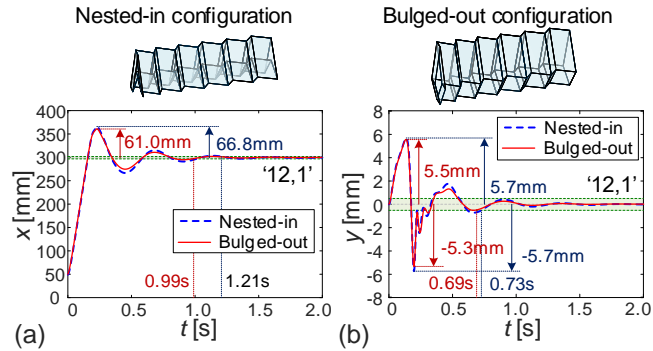

**Fig. 12.** Displacement-time histories of the vertex '12,1' in a nested-in tube and a bulged-out tube under free deployment in (a) the $x$-direction and (b) the $y$-direction. The settling times and the overshoot values are denoted.

## V. SUMMARY AND CONCLUSIONS

With infinite design space, excellent deformability, and extraordinary folding-induced mechanical properties, origami designs have received extensive attention in the development of deployable structures. However, comparing with the enormous progresses achieved in the areas of crease pattern designs, the packaging techniques, the stowed and deployed configurations, and the quasi-static properties, the dynamics of origami deployable structures, particularly, the transient deploying dynamics, has not been well understood due to the lack of a mature modeling methodology. Aiming at advancing the state of the art, this paper investigates the transient dynamic behavior of a Miura-origami tube during free deployment. The reason for choosing the Miura-origami tube as the object of study is because the Miura-origami is relatively simple in geometry design, while the deployable tubes are characterized by strengths in load carrying and shape transforming.

To tackle the transient dynamics problem, a preliminary free-deployment test is firstly carried out on a practical Miura-origami tube prototype. It reveals that in addition to the transient oscillations in the deploying direction, the tube could also exhibit significant transverse vibrations, which are induced by the non-negligible deformations of the creases and the folding deviations between adjacent half-cells. To describe such phenomena, the rigid-folding SDOF dynamic model is no longer adequate. Observations from the experiments provide us with useful guidelines in simplifying dynamic modeling. By admitting the rigid-folding kinematics within each half-cell and replacing the kinematic constraints between adjacent half-cell with additional potential energy, an accurate and processible MDOF dynamic model is developed via the first principles. Based on this model, we are then able to predict the transient dynamics of the tube in both the deploying and the transverse directions.

This paper then carries out a comprehensive study to uncover the effects of the physical and geometric parameters on the transient dynamics, characterized by the settling times and the overshoot values in both the deploying and the transverse directions. An interesting finding is that the parametric dependence relationships are sometimes contradictory in the deploying and the transverse directions. For example, enlarging the damping coefficient is generally favorable in suppressing the transverse vibrations; nevertheless, in the deploying direction, an overlarge damping coefficient would significantly increase the settling time for the tube to be deployed to the desired configuration. Such a contradiction is also encountered when examining the geometries of the origami tube. In the deploying and the transverse directions, the relationships between the overshoot values and the sector angle ($\gamma_A$) are opposite in the overall trend. The obtained relationships between the parameters and the transient dynamic behaviors of the tube lay a solid foundation for developing Miura-origami deployable tubes with robust dynamic performance. Especially, the observed contradictions between the deploying and the transverse directions suggest that a compromise in design is sometimes necessary.

We also want to remark here that the underlying philosophies employed in this research, i.e., extracting and abstracting core features of folding from experiments, making reasonable assumptions and simplifications, are general and important. Although only the Miura-origami tube is exemplified, we believe that the proposed dynamic modeling technique could be applied to other origami structures. This brings up several interesting questions that are worthy of future study. For example, how to conclude a generic dynamic modeling methodology for origami structures with non-ideal creases or facets? How to determine the geometric and physical parameters of a practical origami structure? Without accurately obtaining the parameter values, quantitative predictions of the origami dynamics are impossible. Adopting model-based parameter identification might be a feasible way [58].

# ACKNOWLEDGMENTS

This research was supported by the National Natural Science Foundation of China under award No. 11902078 and No. 11932015, and the Major Research Plan of the National Natural Science Foundation of China under grant no. 91748203.

# Supplementary Information for

# Transient Dynamics of a Miura-Origami Tube during Free Deployment

Haiping Wu[1], Hongbin Fang[2,3,4,*], Lifen Chen[1], and Jian Xu[2,3,4]

[1] Department of Aeronautics and Astronautics, Fudan University, Shanghai 200433, China
[2] Institute of AI and Robotics, Fudan University, Shanghai 200433, China
[3] Engineering Research Center of AI & Robotics, Ministry of Education, Fudan University, Shanghai 20043, China
[4] Shanghai Engineering Research Center of AI & Robotics, Fudan University, Shanghai 200433, China

*Author for correspondence: fanghongbin@fudan.edu.cn (H. Fang)

## S1. Detailed expressions of the nonconservative generalized force

In each half-cell, viscous damping applies on 12 creases, as illustrated in Fig. S1. Based on the principle of virtual work, the nonconservative generalized force can be written as

$$Q_i = \sum_{r=1}^{12}\left(-c_0 l_{i,r}\frac{\mathrm{d}\psi_{i,r}}{\mathrm{d}t}\right)\frac{\partial \psi_{i,r}}{\partial \theta_{i,\mathrm{A}}}, \tag{S1}$$

Where $i = 1, ..., 2N$ denotes the number of half-cells, $r = 1, ..., 12$ indicates the 12 creases, $l_{i,r}$ denotes the length of the $r$-th crease, and $\psi_{i,r}$ are the corresponding dihedral angles. These angles can be expressed as

$$\psi_{i,1} = \varphi_{i,A1}, \ \psi_{i,2} = \psi_{i,3} = \varphi_{i,C}, \ \psi_{i,4} = \varphi_{i,B1},$$

$$\psi_{i,5} = \frac{\varphi_{i,A2} + \varphi_{(i-1),A2}}{2}, \ \psi_{i,6} = \frac{\varphi_{i,A4} + \varphi_{(i-1),A4}}{2},$$

$$\psi_{i,7} = \frac{\varphi_{i,B2} + \varphi_{(i-1),B2}}{2}, \ \psi_{i,8} = \frac{\varphi_{i,B4} + \varphi_{(i-1),B4}}{2}, \quad (S2)$$

$$\psi_{i,9} = \frac{\varphi_{i,A4} + \varphi_{(i+1),A4}}{2}, \ \psi_{i,10} = \frac{\varphi_{i,A2} + \varphi_{(i+1),A2}}{2},$$

$$\psi_{i,11} = \frac{\varphi_{i,B4} + \varphi_{(i+1),B4}}{2}, \ \psi_{i,12} = \frac{\varphi_{i,B2} + \varphi_{(i+1),B2}}{2}.$$

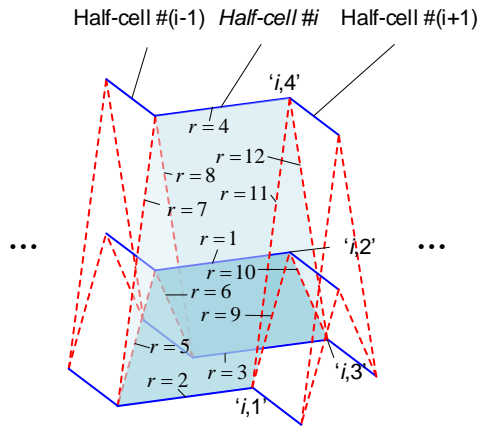

**Fig. S1.** The 12 creases, numbered by $r$ from 1 to 12, in the $i$-th half-cell.

To apply the principle of virtual work, corresponding to $\theta_{i,A}$, except the $i$-th half-cell, deformations of other half-cells are assumed to be zero. Hence, we have,

$$\frac{\mathrm{d}\psi_{i,r}}{\mathrm{d}t} = \frac{\partial \psi_{i,r}}{\partial \theta_{i,A}} \dot{\theta}_{i,A}. \quad (S3)$$

Since the dihedral angles $\varphi_{i,kj}$ ($i=1,...,2N; k=A,B; j=1,...,4$) and $\varphi_{i,C}$ are functions of the folding angle $\theta_{i,A}$, the partial derivatives can be expressed as

$$\frac{\partial \varphi_{i,k2}}{\partial \theta_{i,A}} = \frac{\partial \varphi_{i,k2}}{\partial \theta_{i,k}} \frac{\partial \theta_{i,k}}{\partial \theta_{i,A}}, \ \frac{\partial \varphi_{i,k4}}{\partial \theta_{i,A}} = -\frac{\partial \varphi_{i,k2}}{\partial \theta_{i,A}},$$

$$\frac{\partial \varphi_{i,k1}}{\partial \theta_{i,A}} = -2\frac{\partial \theta_{i,k}}{\partial \theta_{i,A}}, \ \frac{\partial \varphi_{i,C}}{\partial \theta_{i,A}} = \frac{\partial \theta_{i,B}}{\partial \theta_{i,A}} - 1, \ k = A,B. \quad (S4)$$

Substituting Eq. (S2) ~(S4) into Eq.(S1), the nonconservative generalized force corresponding to the generalized coordinate $\theta_{i,A}$ can be obtained,

$$Q_i = -c_0 \dot{\theta}_{i,A} \left( \sum_{k=A,B} \left( 4a_k \left( \pm \frac{\partial \theta_{i,k}}{\partial \theta_{i,A}} \frac{\partial}{\partial \theta_{i,k}} \left( \frac{\varphi_{i,k2}}{2} \right) \right)^2 + b \left( -2\frac{\partial \theta_{i,k}}{\partial \theta_{i,A}} \right)^2 \right) + 2b \left( \frac{\partial \theta_{i,B}}{\partial \theta_{i,A}} - 1 \right)^2 \right). \quad (S5)$$

## S2. Detailed derivations of the equation of motion

The kinetic and potential energies of the tube have been derived In Section III, and the nonconservative generalized force has been expressed in Eq.(S5). By substituting them into the Lagrange equation (i.e., Eq. (7)), the equation of motion can be obtained. Here, we detail the derivation process. Note that the potential energy does not involve the angular velocity terms $\dot{\theta}_{i,A}$, the Lagrange equation can be rewritten as

$$\frac{\mathrm{d}}{\mathrm{d}t}\left(\frac{\partial T}{\partial \dot{\theta}_{i,A}}\right) - \frac{\partial T}{\partial \theta_{i,A}} + \frac{\partial V}{\partial \theta_{i,A}} = Q_i, \quad (S6)$$

The total kinetic energy has been obtained earlier in this paper, i.e.,

$$T = \sum_{i=1}^{2N} \sum_{k=A,B} \sum_{s=x,y,z} T_{i,ks},$$

$$T_{i,kx} = \rho a_k b \sin \gamma_k \left( \frac{1}{3} \dot{L}_i^2 + \dot{x}_{(i-1),1}^2 + \dot{L}_i \dot{x}_{(i-1),1} \right),$$

$$T_{i,ky} = \rho a_k b \sin \gamma_k \left( \frac{1}{3} \dot{S}_i^2 + \frac{4}{3} \dot{W}_i^2 + \dot{y}_{(i-1),1}^2 + 2\dot{W}_i \dot{y}_{(i-1),1} + \dot{S}_i \dot{y}_{(i-1),1} + \dot{W}_i \dot{S}_i \right), \quad (S7)$$

$$T_{i,kz} = \rho a_k b \sin \gamma_k \left( \frac{1}{3} \dot{H}_{i,k}^2 + \dot{z}_{(i-1),1}^2 + \dot{H}_{i,k} \dot{z}_{(i-1),1} \right).$$

The length, width, and height of each half-cell can be expressed by the coordinates of vertices,

$$L_i = x_{i,1} - x_{(i-1),1}, \quad S_i = (-1)^{i-1}\left( y_{i,1} - y_{(i-1),1} \right), \quad z_{(i+1)1} - z_{i1} = 0, \quad (S8)$$

with boundary conditions at the fixed point,

$$x_{0,1} = y_{0,1} = z_{0,1} = 0. \quad (S9)$$

Hence, the kinetic energy $T$ can be expressed as a function of the following derivatives of the external dimensions to time, i.e.,

$$T = T\left( \sum_{i,j=1,\ldots,2N} \left( \dot{L}_i \dot{L}_j, \dot{S}_i \dot{S}_j, \dot{W}_i \dot{S}_j, \dot{W}_i \dot{W}_j, \dot{H}_{i,A} \dot{H}_{j,A}, \dot{H}_{i,B} \dot{H}_{j,B} \right) \right). \quad (S10)$$

If directly substituting Eq.(S10) into Eq.(S6), we will get an extremely complex expression. To simplify the expression, a matrix representation is introduced. To demonstrate the process, only the derivative of the length $L_i$ to the time is exemplified. In detail,

$$\frac{d}{dt}\left( \frac{\partial}{\partial \dot{\theta}_{i,A}} T(\sum_{i,j=1,\ldots,2N} \dot{L}_i \dot{L}_j) \right) - \frac{\partial}{\partial \theta_{i,A}} T(\sum_{i,j=1,\ldots,2N} \dot{L}_i \dot{L}_j) = \frac{d}{dt}\left( \frac{\partial T}{\partial \dot{L}_i} \frac{\partial \dot{L}_i}{\partial \dot{\theta}_{i,A}} \right) - \frac{\partial T}{\partial \dot{L}_i} \frac{\partial \dot{L}_i}{\partial \theta_{i,A}} \quad (S11)$$

Equation (S 11) can be simplified based on the following expressions

$$\dot{L}_i = \frac{\partial L_i}{\partial \theta_{i,A}} \dot{\theta}_{i,A}, \quad \frac{\partial \dot{L}_i}{\partial \dot{\theta}_{i,A}} = \frac{\partial L_i}{\partial \theta_{i,A}}, \quad T_{LL}(i,j) = \frac{\partial^2 T}{\partial \dot{L}_i \partial \dot{L}_j}, \quad \frac{\partial T}{\partial \dot{L}_i} = \sum_{j=1}^{2N} T_{LL}(i,j) \dot{L}_j, \quad (S12)$$

where $T_{LL}(i,j)$ could constitute a constant $2N \times 2N$ matrix $\mathbf{T}_{LL}$. Based on Eq.(S12), Eq. (S11) can be simplified into

$$\frac{\mathrm{d}}{\mathrm{d}t}\left(\frac{\partial T}{\partial \dot{L}_i}\frac{\partial \dot{L}_i}{\partial \dot{\theta}_{i,\mathrm{A}}}\right) - \frac{\partial T}{\partial \dot{L}_i}\frac{\partial \dot{L}_i}{\partial \theta_{i,\mathrm{A}}} = \sum_{j=1}^{2N}G_L(i,j)\dot{\theta}_{j,\mathrm{A}}^2 + \sum_{j=1}^{2N}J_L(i,j)\ddot{\theta}_{j,\mathrm{A}},$$

$$G_L(i,j) = mT_{LL}(i,j)\frac{\partial^2 L_j}{\partial \theta_{j,\mathrm{A}}^2}\frac{\partial L_i}{\partial \theta_{i,\mathrm{A}}}, \quad (S13)$$

$$J_L(i,j) = mT_{LL}(i,j)\frac{\partial L_j}{\partial \theta_{j,\mathrm{A}}}\frac{\partial L_i}{\partial \theta_{i,\mathrm{A}}}.$$

In Eq. (S13), $m_\mathrm{A}$ and $m_\mathrm{B}$ denote the facet mass of the bottom sheet and the top cell, respectively,

$$m = m_\mathrm{A} + m_\mathrm{B},\ m_\mathrm{A} = \rho a_\mathrm{A} b \sin\gamma_\mathrm{A},\ m_\mathrm{B} = \rho a_\mathrm{B} b \sin\gamma_\mathrm{B}. \quad (S14)$$

$G_L(i,j)$ and $J_L(i,j)$ constitute two $2N \times 2N$ matrices $\mathbf{G}_L$ and $\mathbf{J}_L$. Employing such matrix representations and based on the same process, the rest components of the kinetic energy (i.e., $T$ in Eq. (S10)) can be similarly expressed. Substituting all the components into Eq. (S11), we have

$$\frac{\mathrm{d}}{\mathrm{d}t}\left(\frac{\partial T}{\partial \dot{\boldsymbol{\theta}}}\right) - \frac{\partial T}{\partial \boldsymbol{\theta}} = \mathbf{J}\ddot{\boldsymbol{\theta}} + \mathbf{G}\dot{\boldsymbol{\theta}}^2,$$

$$\mathbf{J} = \mathbf{J}_L + \mathbf{J}_S + \mathbf{J}_W + \mathbf{J}_{WS} + \mathbf{J}_{H_\mathrm{A}} + \mathbf{J}_{H_\mathrm{B}}, \quad (S15)$$

$$\mathbf{G} = \mathbf{G}_L + \mathbf{G}_S + \mathbf{G}_W + \mathbf{G}_{WS} + \mathbf{G}_{H_\mathrm{A}} + \mathbf{G}_{H_\mathrm{B}}.$$

The total potential energy has also been derived earlier, i.e.,

$$V = V_b + V_S + V_\theta.$$

$$V_b = \sum_{i=1}^{2N}\frac{1}{2}kb\left[(\varphi_{i,\mathrm{A1}} - \varphi_{i,\mathrm{A1}}^0)^2 + (\varphi_{i,\mathrm{B1}} - \varphi_{i,\mathrm{B1}}^0)^2 + 2(\varphi_{i,\mathrm{C}} - \varphi_{i,\mathrm{C}}^0)^2\right],$$

$$V_S = 2\sum_{i=1}^{2N-1}\frac{1}{2}ka_\mathrm{A}\left(\frac{\varphi_{i,\mathrm{A2}} + \varphi_{(i+1),\mathrm{A2}}}{2} - \frac{\varphi_{i,\mathrm{A2}}^0 + \varphi_{(i+1),\mathrm{A2}}^0}{2}\right)^2 \quad (S16)$$

$$+ 2\sum_{i=1}^{2N-1}\frac{1}{2}\mu ka_\mathrm{B}\left(\frac{\varphi_{i,\mathrm{B2}} + \varphi_{(i+1),\mathrm{B2}}}{2} - \frac{\varphi_{i,\mathrm{B2}}^0 + \varphi_{(i+1),\mathrm{B2}}^0}{2}\right)^2,$$

$$V_\theta = 2\sum_{i=1}^{2N-1}\frac{1}{2}k^*\left[(\xi_{i,\mathrm{A}} - \xi_{(i+1),\mathrm{A}})^2 + (\xi_{i,\mathrm{B}} - \xi_{(i+1),\mathrm{B}})^2\right].$$

Hence, the derivatives of $V_b$ and $V_\theta$ to the time can be expressed as

$$\frac{\partial V_b}{\partial \theta_{i,\mathrm{A}}} = kb\left[\sum_{k=\mathrm{A,B}}\left((\varphi_{i,k1} - \varphi_{i,k1}^0)\frac{\partial \varphi_{i,k1}}{\partial \theta_{i,\mathrm{A}}}\right) + 2(\varphi_{i,\mathrm{C}} - \varphi_{i,\mathrm{C}}^0)\frac{\partial \varphi_{i,\mathrm{C}}}{\partial \theta_{i,\mathrm{A}}}\right],$$

$$\frac{\partial V_\theta}{\partial \theta_{i,\mathrm{A}}} = 2k^*\sum_{j=1}^{2N-1}\sum_{k=\mathrm{A,B}}(\xi_{j,k} - \xi_{(j+1),k})\frac{\partial \xi_{j,k}}{\partial \theta_{i,\mathrm{A}}}. \quad (S17)$$

To derive the derivative of $V_S$ with respect to $\theta_{i,A}$, we first examine $\partial V_S / \partial \varphi_{i,k2}$, which can be expressed as

$$\frac{\partial V_S}{\partial \varphi_{i,k2}} = \sum_{j=1}^{2N} P_\varphi(i,j)\varphi_{j,k2}, \quad P_\varphi(i,j) = \frac{\partial^2 V_S}{\partial \varphi_{i,k2} \partial \varphi_{j,k2}},$$

$$\frac{\partial V_S}{\partial \varphi_{i,k2}} = \sum_{j=1}^{2N} P_{\varphi k}(i,j)\varphi_{j,k2}, \quad P_{\varphi k}(i,j) = \frac{\partial^2 V_S}{\partial \varphi_{i,k2} \partial \varphi_{j,k2}}, k = A, B \quad (S18)$$

where $P_{\varphi k}(i,j)$ constitutes two constant $2N \times 2N$ matrixes, namely, $P_{\varphi A}(i,j)$ and $P_{\varphi B}(i,j)$; they satisfy $\mathbf{P}_{\varphi B} = k a_B / a_A \mathbf{P}_{\varphi A}$. Based on this equation, we have

$$\frac{\partial V_S}{\partial \theta_{i,A}} = \sum_{k=A,B} \sum_{j=1}^{2N} P_{\varphi k}(i,j)\varphi_{j,k2} \frac{\partial \varphi_{i,k2}}{\partial \theta_{i,k}} \frac{\partial \theta_{i,k}}{\partial \theta_{i,A}}. \quad (S19)$$

Based on Eq. (17) and (19), we can use a vector $\mathbf{F}_V$ to express $\partial V / \partial \mathbf{\theta}$,

$$\mathbf{F}_V = \frac{\partial V}{\partial \mathbf{\theta}} = \frac{\partial V_S}{\partial \mathbf{\theta}} + \frac{\partial V_b}{\partial \mathbf{\theta}} + \frac{\partial V_\theta}{\partial \mathbf{\theta}}. \quad (S20)$$

Substituting Eq.(S5), Eq.(S15), and Eq.(S20) into Eq.(S6), the equation of motion of the Miura-ori tube can be obtained, which can be expressed as the following matrix form

$$\mathbf{J}\ddot{\mathbf{\theta}} + \mathbf{C}\dot{\mathbf{\theta}} + \mathbf{G}\dot{\mathbf{\theta}}^2 + \mathbf{F}_V = \mathbf{0}, \quad (S21)$$

in which, $\mathbf{C}$ is a diagonal matrix, whose elements are

$$\mathbf{C}(i,i) = -\frac{Q_i}{\dot{\theta}_{i,A}}. \quad (S22)$$